# MiDeSeC: A Dataset for Mitosis Detection and Segmentation in Breast Cancer Histopathology Images

**Organizers**

• Department of Computer Engineering, Ankara University, Ankara, Turkey

   – Prof. Dr. Refik Samet, samet@eng.ankara.edu.tr
   – PhD Student Zeynep Yildirim, yildirimz@ankara.edu.tr
   – PhD Student Nooshin Nemati, nntolakan@ankara.edu.tr
   – MSc Student Mohamed Traore, mtraore@ankara.edu.tr

• Department of Software Engineering, Mehmet Akif Ersoy University, Burdur, Turkey

   – Assoc. Prof. Dr. Emrah Hancer, ehancer@mehmetakif.edu.tr

• Department of Medical Pathology, Ankara University, Ankara, Turkey

   – Prof. Dr. Serpil Sak, sak@medicine.ankara.edu.tr
   – Assoc. Prof. Dr. Bilge Ayca Kirmizi, akarabork@yahoo.com

## 1 Introduction

Nottingham Grading System [1] emphasizes three key morphological features on Hematoxylin and Eosin (H&E) stained slides to grade breast cancer: mitotic count, tubule formation, and nuclear pleomorphism. Mitotic count is the most prominent feature among them. Searching for mitosis on glass slides is a routine procedure for breast pathologists. Since there are so many high power fields (HPFs) on a single slide and mitotic cells vary in appearance, it is a tedious and time-consuming task. Additionally, mitotic cell judgment is somewhat subjective, making it difficult for pathologists to reach a consensus. Thus, it is extremely important to develop automatic detection methods that will not only save time and material resources, but will also enhance the reliability of pathological diagnosis.

It is difficult to detect mitosis in H&E stained images because of the following challenges. First of all, mitosis appears in many forms. There are four stages in mitosis: prophase, metaphase, anaphase, and telophase. Cells in different stages have different shapes and structures. The telophase stage occurs when a cell's nucleus has split into two parts, but the nuclei are still considered one cell since they are not totally separated. It is also noteworthy that mitotic cells are significantly less numerous than non-mitotic cells. As a result of their low probability, their detection is more complicated. Third, some cells (e.g. apoptotic cells and dense nuclei) have similar appearances to mitosis, which makes them challenging to detect. In recent years, many methods for automatic mitosis detection have been proposed. Some mitosis detection contests contributed to this phenomenon, including the 2012 ICPR mitosis detection contest [2], the AMIDA13 contest at MICCAI

2013 [3], the 2014 ICPR MITOS-ATYPIA challenge [4], and the TUPAC16 contest at MICCAI 2016 [5]. However, these contests except for TUPAC16 did not specifically focus on breast cancer, the second most frequent cause of death after lung cancer in terms of cancer deaths among women. This study introduces a publicly available dataset (called MiDeSeC) for the detection and segmentation process of mitosis in H&E stained breast cancer histopathology images. The MiDeSeC dataset is therefore expected to aid in developing robust and reliable cancer grading systems.

## 2 Dataset

The MiDeSeC dataset is created through H&E stained invasive breast carcinoma, no special type (NST) slides of 25 different patients captured at 40x magnification from the Department of Medical Pathology at Ankara University. The slides have been scanned by 3D Histech Panoramic p250 Flash-3 scanner and Olympus BX50 microscope. As several possible mitosis shapes exist, it is crucial to have a large dataset to cover all the cases. Accordingly, a total of 50 regions is selected from glass slides for 25 patients, each of regions with a size of 1024×1024 pixels. There are more than 500 mitoses in total in these 50 regions. Two-thirds of the regions are reserved for training, the other third for testing.

There is a ground truth text file (CSV format) included with each HPF, indicating the coordinates of all mitosis regions. The organization of a CSV text file containing mitosis coordinates is as follows:

• The file has no header.

• Each line shows the coordinates of all pixels belonging to the same mitosis.

• Pixels have the following coordinates: x location, y location (Fig. 1).

• The image origin is in the top left corner (0,0) (Fig. 1).

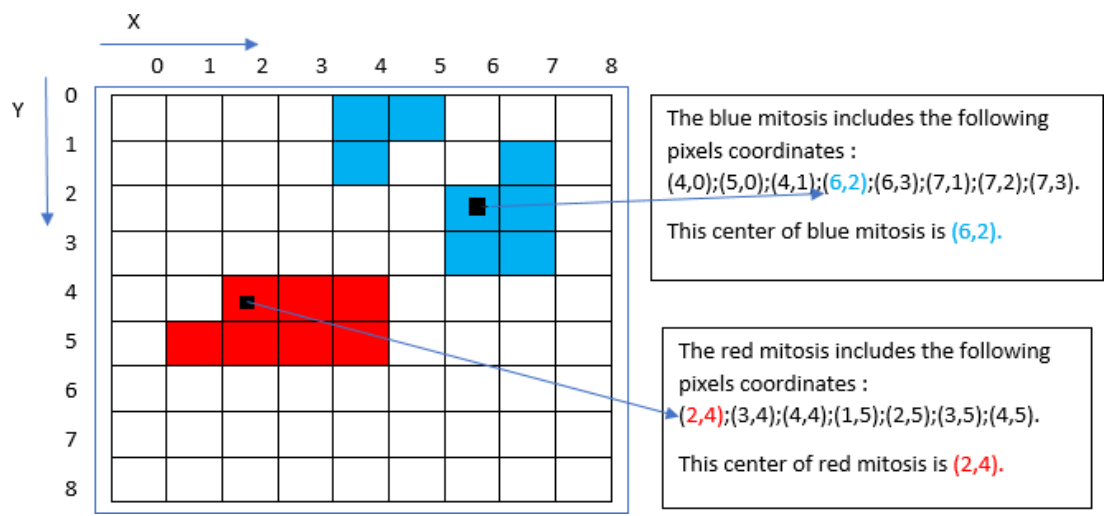

Figure 1: An illustrative example for mitoses representation in a digital image.

We provide an illustrative example in Fig. 1 to show how mitoses are represented in a digital image. An image or graphic display's coordinate system is typically centered in the top-left corner (0, 0). Moreover, pixel coordinates use integer values, i.e., floating-point values are rounded to integers. As an example, a pixel specified as (7.89, 3.02) is displayed at (7, 3). Furthermore, as seen in Fig. 1, while the red structure represents a single mitosis, the two blue structures also represent a single mitosis. In other words, some mitoses may involve gap in their shapes. There is a text file (CSV format) for each image, which contains a list of mitoses.

## 3 Evaluation

To evaluate the goodness of a mitosis detection and segmentation method on the MiDeSeC dataset the defined evaluation metrics are as follows.

$$\text{recall} = \frac{\text{TP}}{\text{TP+FN}}, \qquad (1)$$

$$\text{precision} = \frac{\text{TP}}{\text{TP+FP}}, \qquad (2)$$

$$\text{F1} - \text{Score} = \frac{\text{precision} \times \text{recall}}{\text{precision} + \text{recall}} \qquad (3)$$

where TP (True Positive) is the number of mitoses that are ground truth mitosis among the detected mitoses; FP (False Positive) is the number of mitoses that are not ground truth mitosis among the detected mitoses; and FN (False Negatives) is the number of ground truth mitoses that have not been detected.

**MiDeSeC Dataset Access Link**

The MiDeSeC dataset can be downloaded from the following link:
[Dataset available here](#)

**Acknowledgement**

This work is supported by Turkish Scientific and Research Council (TUBITAK) under Grant No.121E379.